\documentclass[preprint,preprintnumbers,nofootinbib,aps,superscriptaddress,10pt,twocolumn]{revtex4-1}       
\usepackage{amsmath,amssymb,bm,epsfig}
\usepackage{color}
\usepackage{natbib}
\usepackage{hyperref}
\usepackage{ulem}
\usepackage{graphicx}
\usepackage{xcolor,colortbl}
\usepackage{pifont}
\usepackage{float}
\usepackage{dcolumn}
\usepackage{booktabs}
\usepackage{multirow}

\def \beq{\begin{equation}}
\def \eeq{\end{equation}}
\def \beqa{\begin{eqnarray}}
\def \eeqa{\end{eqnarray}}
\def \l{\left(}
\def \r{\right)}

\newcommand{\nn}{\nonumber}

\usepackage{xspace}

\newcommand{\sNN}{\sqrt{s_{\rm NN}}}

\begin{document}

\title{Role of system size on freezeout conditions extracted from transverse momentum spectra of hadrons}

 \author{Ajay Kumar Dash}
 \email{ajayd@niser.ac.in}
 \affiliation{School of Physical Sciences, National Institute 
 of Science Education and Research, HBNI, 
 Jatni, 752050, India}
 
 \author{Ranbir Singh}
 \email{ranbir.singh@niser.ac.in}
 \affiliation{School of Physical Sciences, National Institute 
 of Science Education and Research, HBNI, 
 Jatni, 752050, India}

 \author{Sandeep Chatterjee}
 \email{Sandeep.Chatterjee@fis.agh.edu.pl}
 \affiliation{AGH University of Science and Technology,
 Faculty of Physics and Applied Computer Science,
 al. Mickiewicza 30, 30-059 Krakow, Poland}
 
 \author{Chitrasen Jena}
 \email{cjena@iisertirupati.ac.in}
 \affiliation{Indian Institute of Science Education and Research, 
 Tirupati, 517507, India}
   
 \author{Bedangadas Mohanty}
 \email{bedanga@niser.ac.in}
 \affiliation{School of Physical Sciences, National Institute 
 of Science Education and Research, HBNI, 
 Jatni, 752050, India}
 
\begin{abstract}
The data on hadron transverse momentum spectra in different centrality classes of p+Pb collisions 
at $\sNN=5.02$ TeV has been analysed to extract the freezeout hypersurface within a simultaneous 
chemical and kinetic freezeout scenario. The freezeout hypersurface has been extracted for three 
different freezeout schemes that differ in the way strangeness is treated: i. unified freezeout 
for all hadrons in complete thermal equilibrium (1FO), ii. unified freezeout for all hadrons with 
an additional parameter $\gamma_S$ which accounts for possible out-of-equilibrium production of 
strangeness (1FO$+\gamma_S$), and iii. separate freezeout for hadrons with and without strangeness 
content (2FO). Unlike in heavy ion collisions where 2FO performs best in describing the mean hadron 
yields as well as the transverse momentum spectra, in p+Pb we find that 1FO$+\gamma_S$ with one less 
parameter than 2FO performs better. This confirms expectations from previous analysis on the system 
size dependence in the freezeout scheme with mean hadron yields: while heavy ion collisions that are 
dominated by constituent interactions prefer 2FO, smaller collision systems like proton + nucleus 
and proton + proton collisions with lesser constituent interaction prefer a unified freezeout scheme 
with varying degree of strangeness equilibration.

PACS numbers:

\end{abstract}
\maketitle

\section{Introduction}\label{sec.intro}

The knowledge of the surface of last scattering of the hadrons produced in a heavy ion 
collision (HIC) event is of utmost significance as it contributes to the calibration of 
the hadronic physics baseline to be contrasted with data to extract information of the 
quark gluon plasma (QGP) phase \cite{Gyulassy:2004zy, Adams:2005dq} as well as those of the QCD critical point \cite{Rajagopal:2000wf, Stephanov:2004wx}. The hadron 
resonance gas model has been the main phenomenological model to extract the freezeout 
hypersurface by comparing to the data of hadron yields~\cite{BraunMunzinger:1995bp, Yen:1998pa, 
Cleymans:1999st, Becattini:2000jw, Andronic:2005yp, Chatterjee:2015fua} as well as 
spectra~\cite{Broniowski:2001we, Broniowski:2001uk, Begun:2013nga, Chatterjee:2014lfa}. 
The surface where the hadrons cease to interact inelastically is known as the chemical 
freezeout surface. The hadron yields freeze here. The surface where the hadrons cease to 
interact even elastically is known as the kinetic freezeout surface. 
The shape of the transverse momentum spectra of hadrons get fixed here. 
Depending on the model assumptions, the chemical and kinteic freezeout surfaces could be separate~\cite{Schnedermann:1993ws, 
Afanasiev:2002fk, BurwardHoy:2002xu} or together~\cite{Letessier:1994cn, Csorgo:1994dd, Csernai:1995zn, 
Broniowski:2001we, Broniowski:2001uk}. In this study, we have worked with the THERMINATOR event generator 
where a combined frezeout of both yields as well as spectra at the same surface is 
implemented~\cite{Kisiel:2005hn, Chojnacki:2011hb}.

Traditionally, a single unified freezeout of all hadrons have been studied (1FO)~\cite{Cleymans:1999st, 
Becattini:2000jw, Andronic:2005yp}. However, the data from LHC have thrown open the interpretation of 
freezeout and several alternate schemes have been proposed~\cite{Steinheimer:2012rd, Becattini:2012xb, 
Petran:2013lja, Bellwied:2013cta, Chatterjee:2013yga, Bugaev:2013sfa, Floris:2014pta, Noronha-Hostler:2014usa, 
Bazavov:2014xya, Alba:2016hwx}. In the standard picture, freezeout is interpreted as a competition 
between fireball expansion and interaction of the constituents. Thus it is natural to expect 
system size dependence in freezeout conditions, since constituent interactions decrease as one goes 
from nucleus-nucleus (A+A) to proton-nucleus (p+A) and proton-proton (p+p) collisions. On the 
contrary, it was found that 1FO provides equally good description of data on mean hadron yields 
of $e^++e^-$, p+p and A+A~\cite{Becattini:2010sk}. This lack of sensitivity of the 1FO approach on 
the varying rate of interaction amongst the constituents and fireball expansion across system size raises 
doubt on the standard interpretation of freezeout as a competition between expansion and interaction.

In Ref.~\cite{Chatterjee:2016cog}, the yield data was analysed within three different approaches: i. 1FO, 
ii. single unified freezeout of all hadrons with an additional parameter $\gamma_S$ accounting for non-equilibrium 
production of strangeness (1FO+$\gamma_S$), and iii. separate freezeout surface for hadrons with and 
without strangeness content (2FO). The data on hadron yield was analysed across systems: p+p, p+Pb 
and Pb+Pb enabling one to study the freezeout condition for mid-rapidity charged particle multiplicity 
as well as the system volume varying over three orders of magnitude. It was found that while 1FO and 
1FO+$\gamma_S$ schemes are blind to system size, 2FO exhibits a strong system size dependence. While 
for central and mid-central collisions, 2FO provides the least chi-square per degree of freedom, for 
peripheral Pb+Pb to all centralities of p+Pb and min bias p+p, 1FO$+\gamma_S$ provides a better 
description. This emphasizes a plausible freezeout scenario: in case of large system sizes, the freezeout 
dynamics is dominated by hadron interactions and hence flavor dependence in hadron-hadron cross sections 
play a role resulting in 2FO being the preferred freezeout scheme. On the other hand, in small systems 
the freezeout is mostly driven by rapid expansion and little interaction resulting in a sudden and 
rapid freezeout and hence disfavoring 2FO. 

In this paper, we extend the above line of argument by studying the data on hadron spectra. The 2FO prescription has 
been already demonstrated to describe better the data on hadron spectra than 1FO in Pb+Pb at $\sNN=2.76$ 
TeV~\cite{Chatterjee:2014lfa}. Here we study the data on hadron spectra in p+Pb at $\sNN=5.02$ TeV \cite{pPbALICEpikp, pPbALICEstrange, pPbALICEresonance} and finally connect to our previous findings with the spectra data in Pb+Pb~\cite{Chatterjee:2014lfa}. 
The spectra of $\pi^{+}+\pi^{-}$, K$^{+}$+K$^{-}$, p+$\bar{\mathrm{p}}$, $\phi$,  $\Lambda$+$\bar{\Lambda}$,  $\Xi$+$\bar{\Xi}$ and $\Omega$+$\bar{\Omega}$ are used for this study which are measured in the mid rapidity 
(0$<$$y_{cm}$$<$0.5) 
by the ALICE collaboration. We have performed the centrality dependence of this study by analyzing the data in seven centrality classes, 0-5\%, 5-10\%, 10-20\%, 20-40\%, 40-60\%, 60-80\% and 60-100\%.

The paper is arranged in the following way. In Sec.~ \ref{mod} we discussed about the model used for this study. The results from the model and data are compared in Sec.~\ref{res}. Finally we summarize our findings in Sec.~\ref{sum}.

\section{Model}
\label{mod}
We have studied the data on hadron spectra in 3 schemes: 1FO, 1FO$+\gamma_S$ and 2FO using the THERMINATOR 
event generator~\cite{Kisiel:2005hn, Chojnacki:2011hb}. While 1FO is implemented in the 
standard version of THERMINATOR, in Ref.~\cite{Chatterjee:2014lfa} the standard version of THERMINATOR was 
extended to include the 2FO scheme. We now briefly describe the implementation of the freezeout hypersurface 
and the relevant parameters to be extracted in this approach.

The Cooper-Frye prescription provides the hadron spectra emanating from a freezeout hypersurface
\beqa
 \frac{d^2N}{dyp_Tdp_T} &=& \int d\Sigma\cdot pf\l p\cdot u, T, \gamma_S, \mu\r\label{eq.CF}
\eeqa
where $T$ is the temperature, $\mu=\{\mu_B,\mu_Q,\mu_S\}$ refer to the three chemical potentials 
corresponding to the three conserved charges of QCD, $u^{\mu}$ is the 4-velocity, $d\Sigma^{\mu}$ is the 
differential element of the freezeout hypersurface over which the integration in Eq.~\ref{eq.CF} is supposed 
to be, $p$ is the four momentum. There could be different choices for the parametrization of the freezeout hypersurface and $u^{\mu}$. We 
have worked within the Krakow model \cite{Broniowski:2001we} whereby the freezeout is assumed to occur at a constant proper time $\tau_f$ 
\beqa
\tau_f^2 &=& t^2 - x^2 - y^2 - z^2\label{eq.tauf}
\eeqa
while $u^{\mu}$ is chosen to be
\beqa
 u^\mu &=& x^\mu/\tau_f\label{eq.umu}
\eeqa
where $(t,x,y,z)$ is the space-time cooridinate.\\

THERMINATOR accounts for both primary production as well as secondary contribution from resonance decays when 
evaluating the distribution function $f$. The integration in Eq.~\ref{eq.CF} occurs over the freezeout hypersurface 
coordinates, namely the spacetime rapidity $\eta_s$ whose integration range is from minus infinity to plus infinity, the azimuthal 
angle $\phi$ which is integrated from 0 to $2\pi$ and $\rho=\sqrt{x^2+y^2}$, the perpendicular distance between the $Z$-axis 
and the freezeout hypersurface. $\rho$ is integrated from 0 to $\rho_\text{max}$. Thus, we have 3 parameters within 
the 1FO scheme: $T$, $\tau_f$ and $\rho_\text{max}$ to be extracted by comparison with data.

The choice of the thermodynamic ensemble is a relevant topic whenever one discusses system size dependence. In p+p collisions at the highest SPS and RHIC energies, the use of canonical ensemble or strangeness canonical ensemble has been suggested \cite{Kraus:2007hf, Kraus:2008fh}. At the LHC energies, grand canonical ensemble was found to work best in describing the hadron yields \cite{Das:2016muc}. Similar recent studies on the role of thermodynamic ensemble in small systems can be found in Refs. \cite{Begun:2018qkw, Sharma:2018owb, Sharma:2018uma}. Here, we work with the grand canonical ensemble as well.
Since we work with the LHC data that shows a very good particle-antiparticle symmetry, we have set all the chemical potentials to zero. In 
1FO$+\gamma_S$, there is also the additional parameter $\gamma_S$ in $f$ that accounts for out-of-equilibrium production 
of strangeness. In 2FO, we have different parameter sets for parametrising the non-strange ($T_\text{ns}$, ${\tau_f}_\text{ns}$ 
and ${\rho_\text{max}}_\text{ns}$) and strange ($T_\text{s}$, ${\tau_f}_\text{s}$ and ${\rho_\text{max}}_\text{s}$) freezeout 
hypersurfaces.

\section{Result}
\label{res}
\begin{figure*}
 \begin{center}
 \includegraphics[scale=0.28]{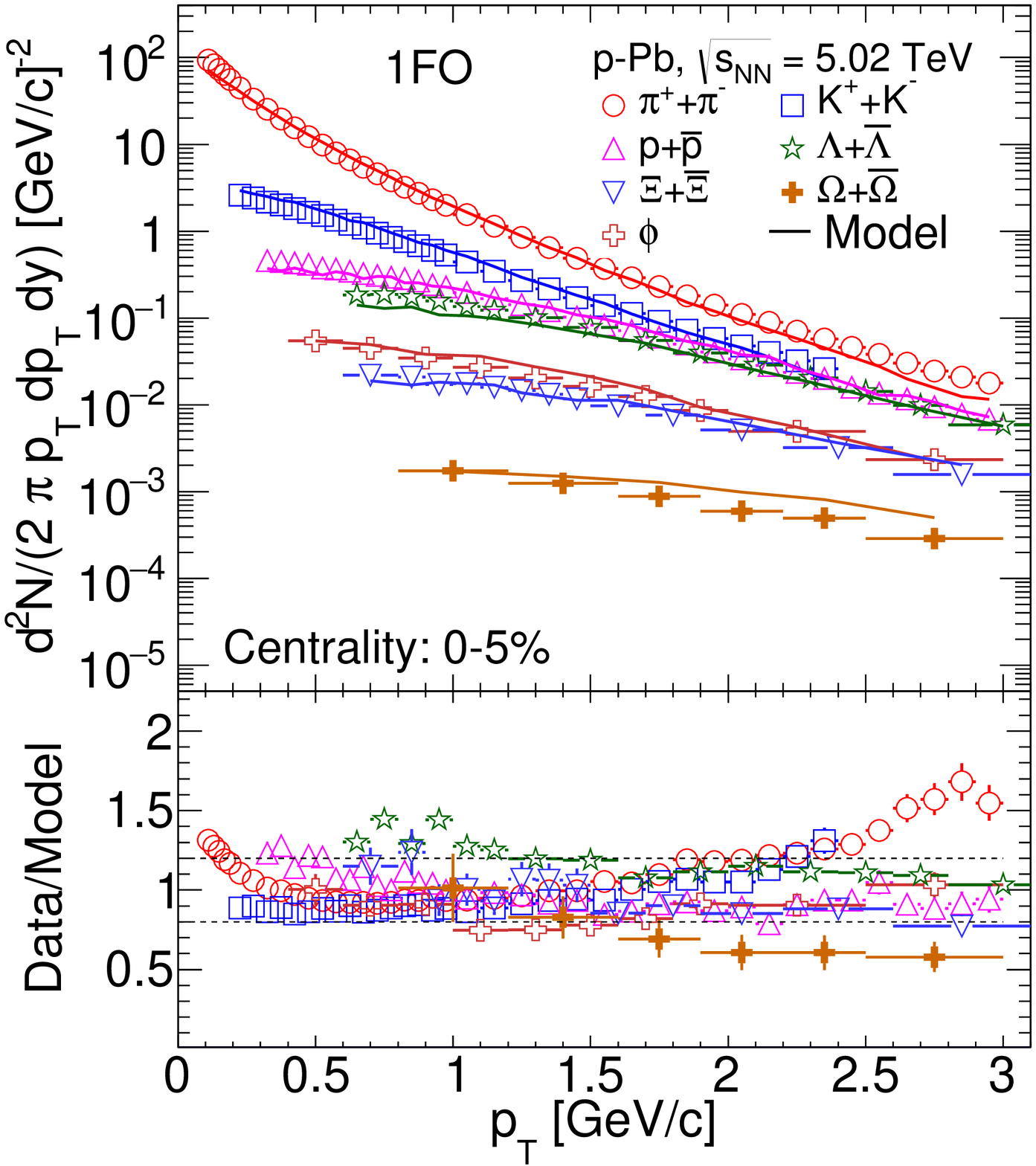}
 \includegraphics[scale=0.28]{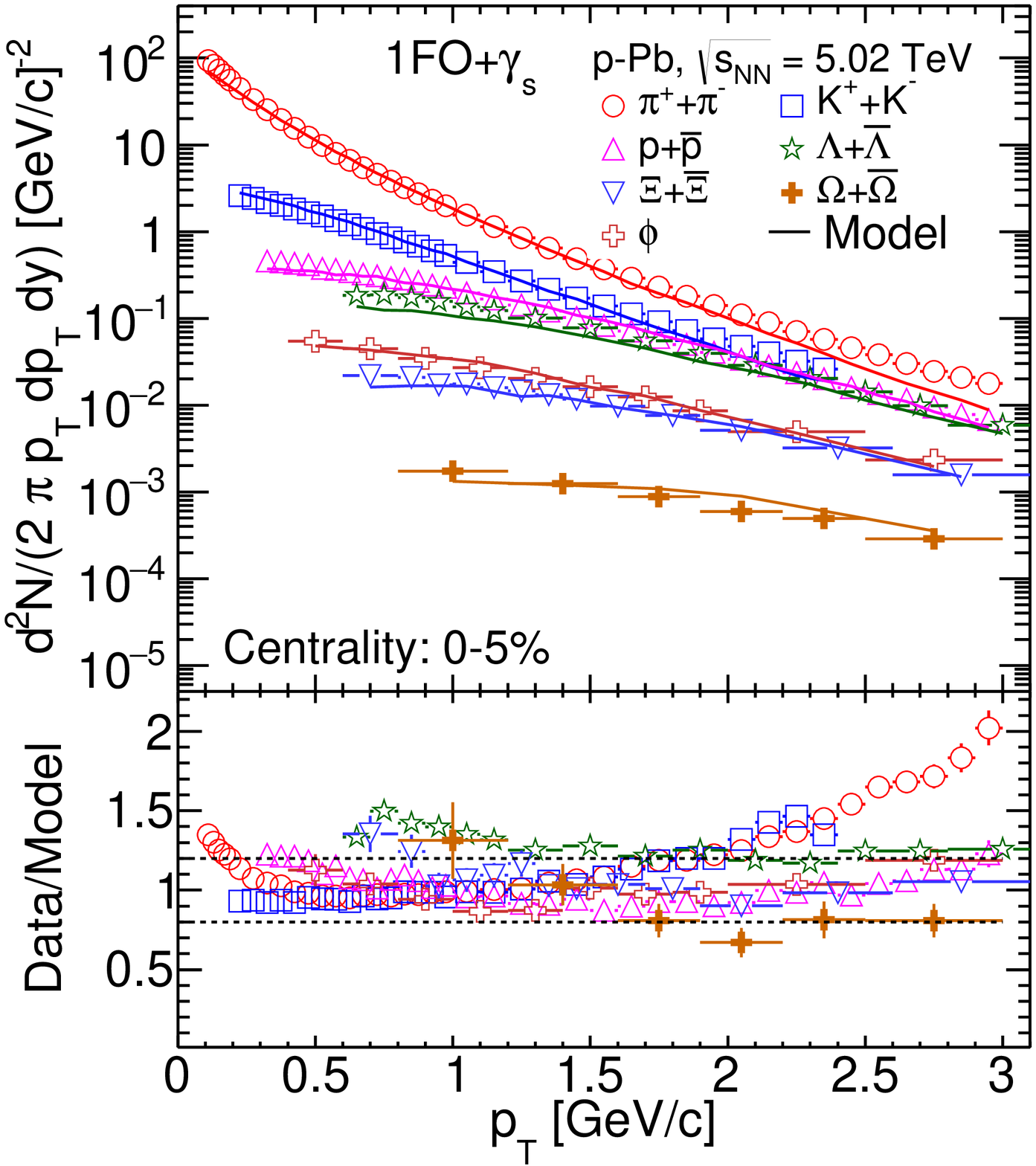} 
 \includegraphics[scale=0.28]{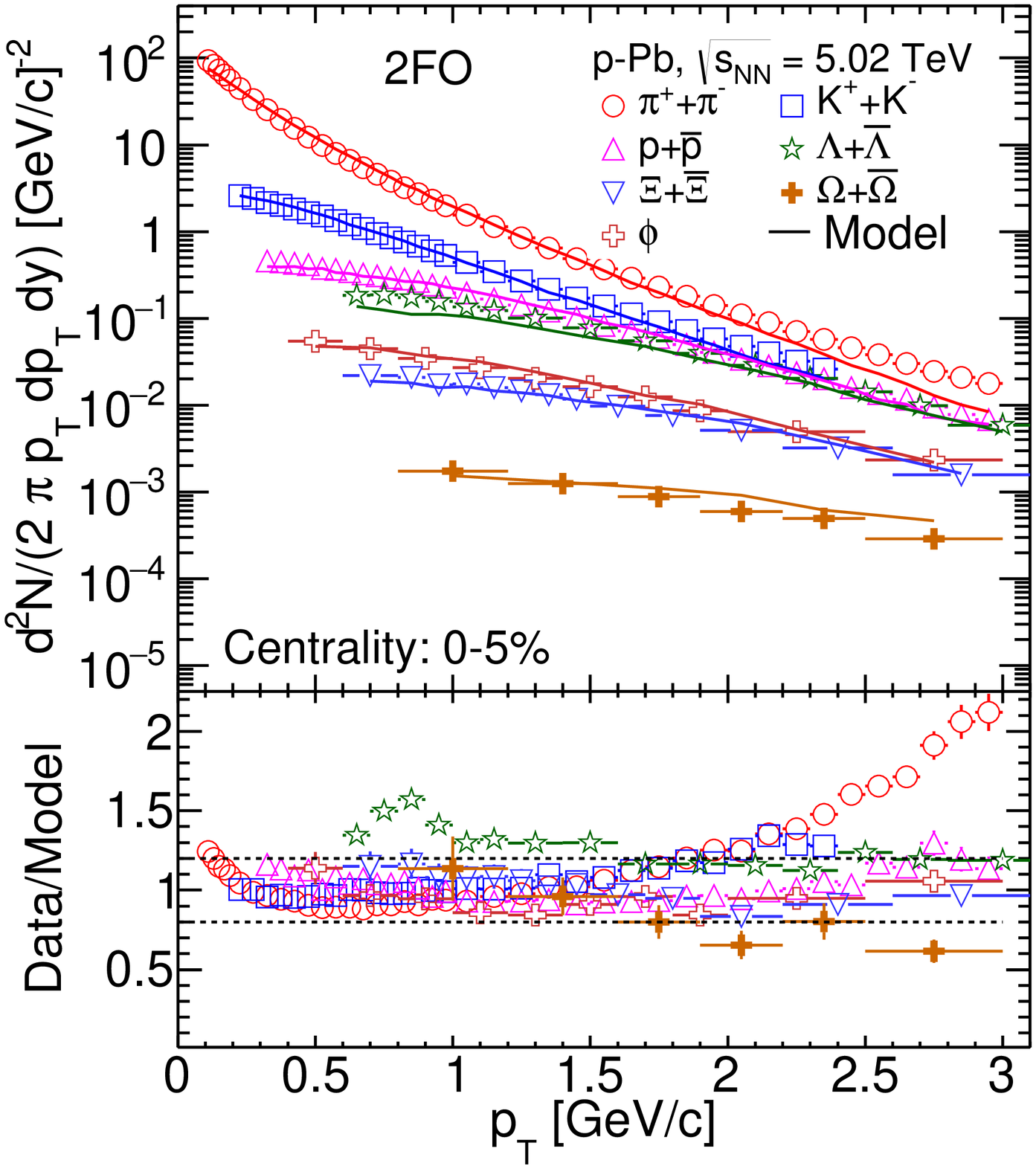}\\
 \includegraphics[scale=0.28]{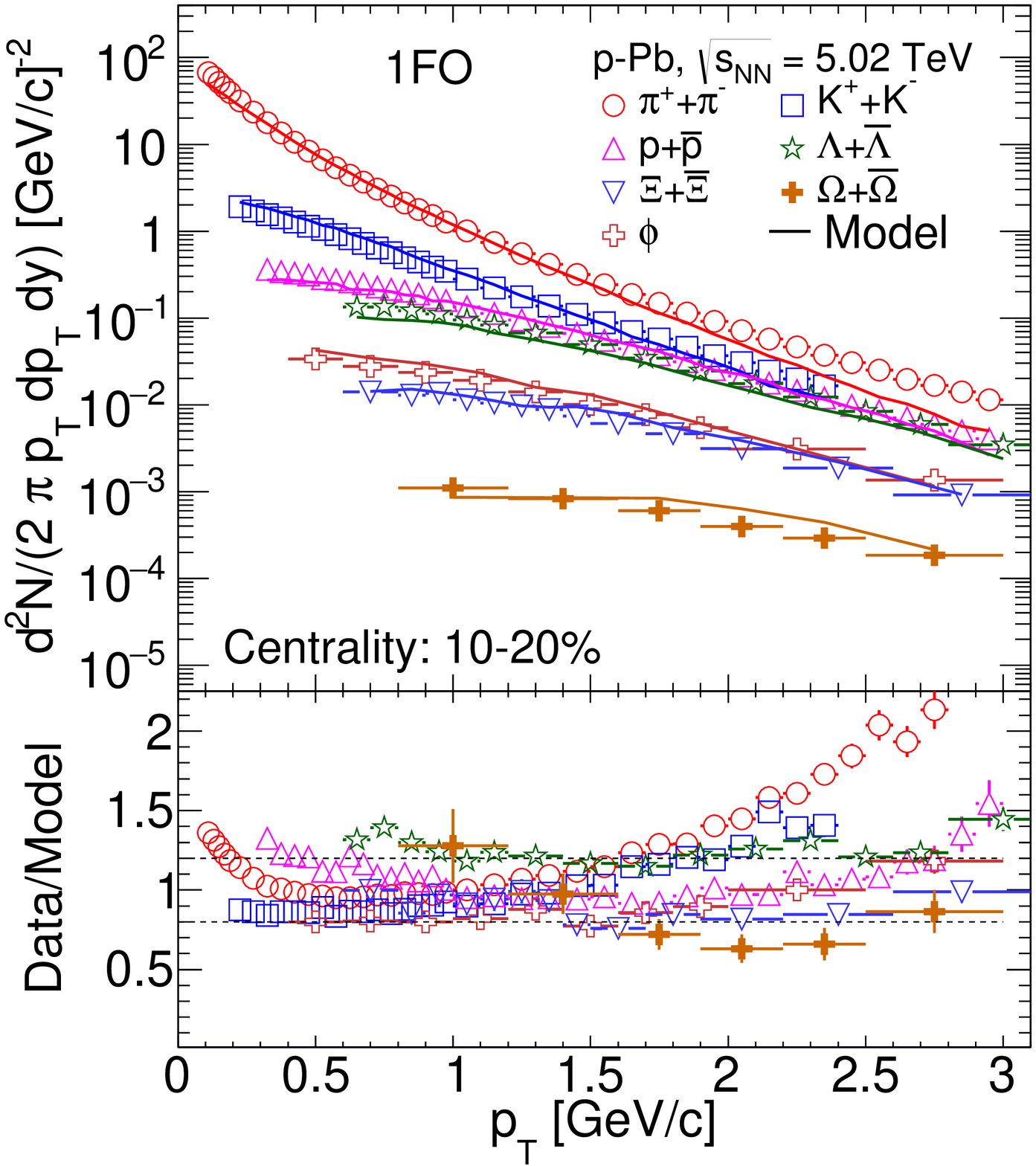}
 \includegraphics[scale=0.28]{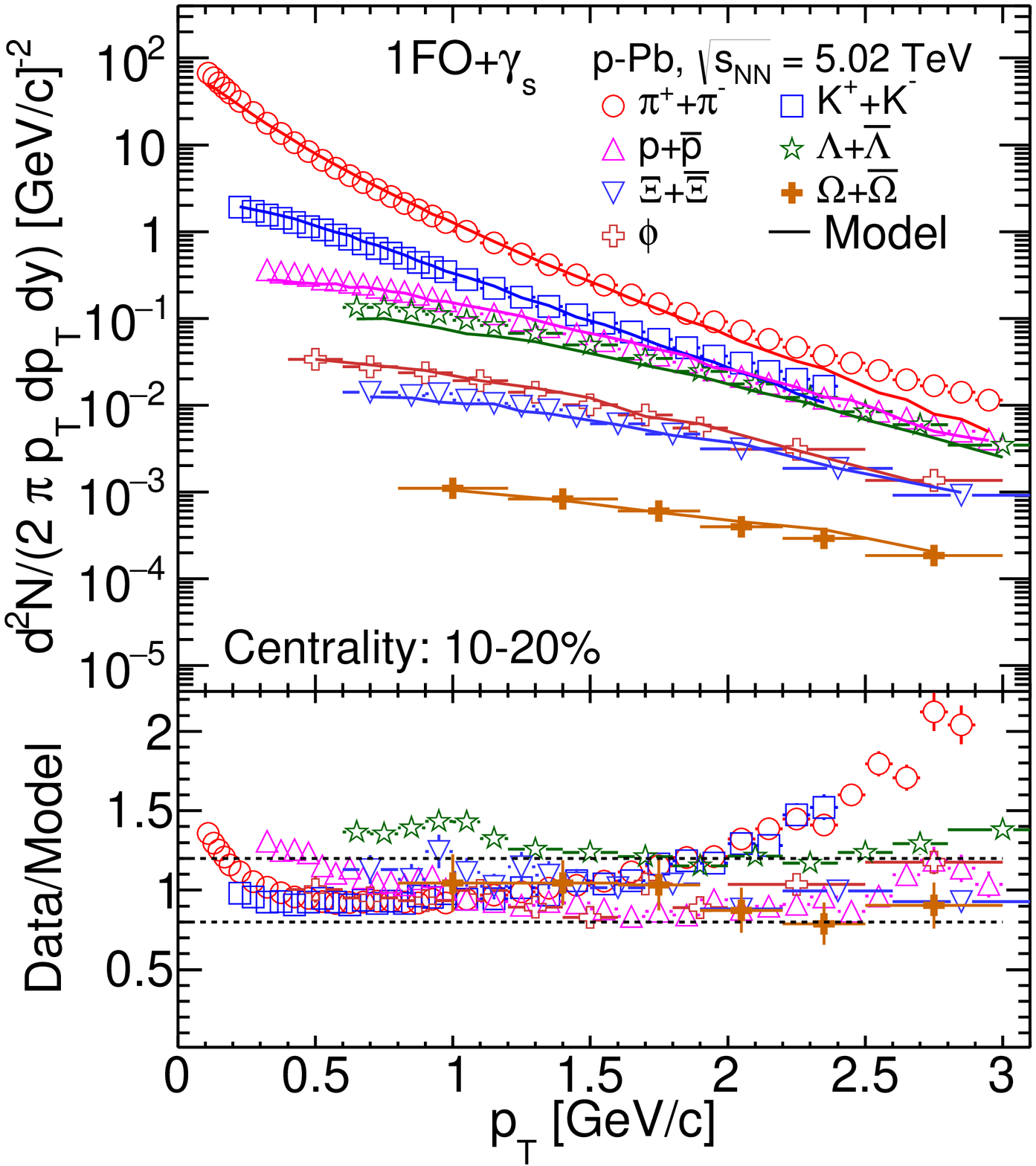} 
 \includegraphics[scale=0.28]{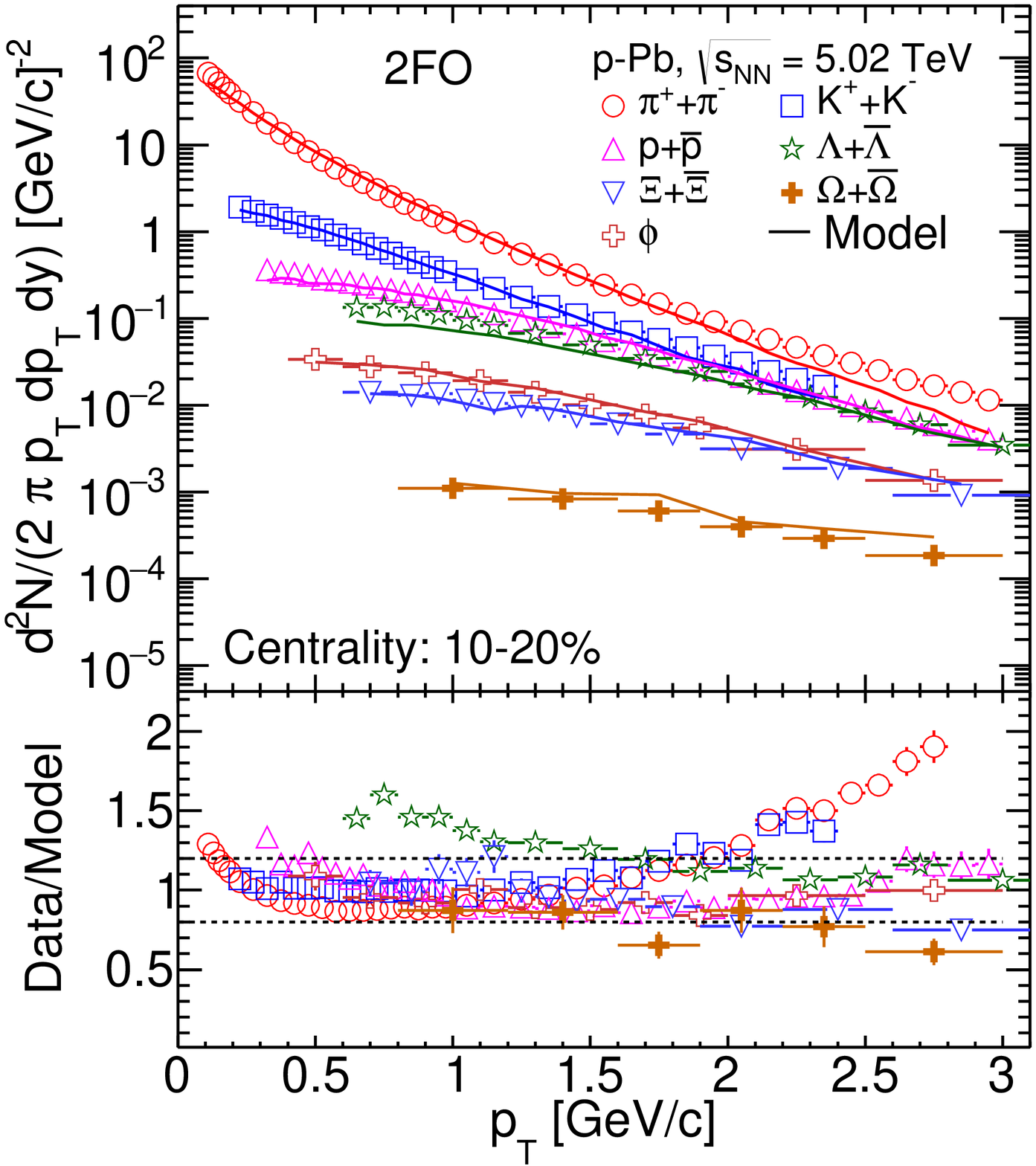}\\
 \includegraphics[scale=0.28]{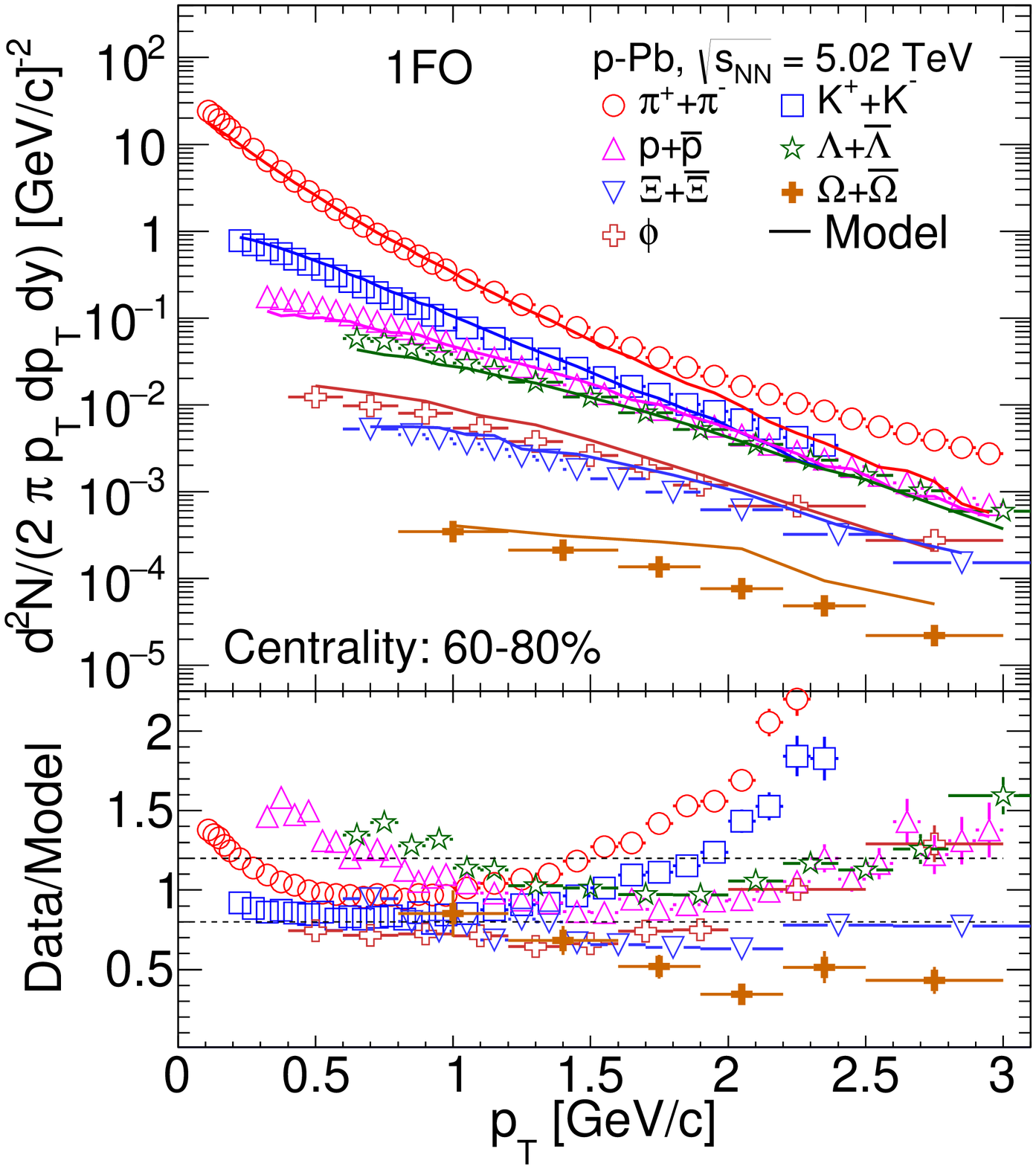}
 \includegraphics[scale=0.28]{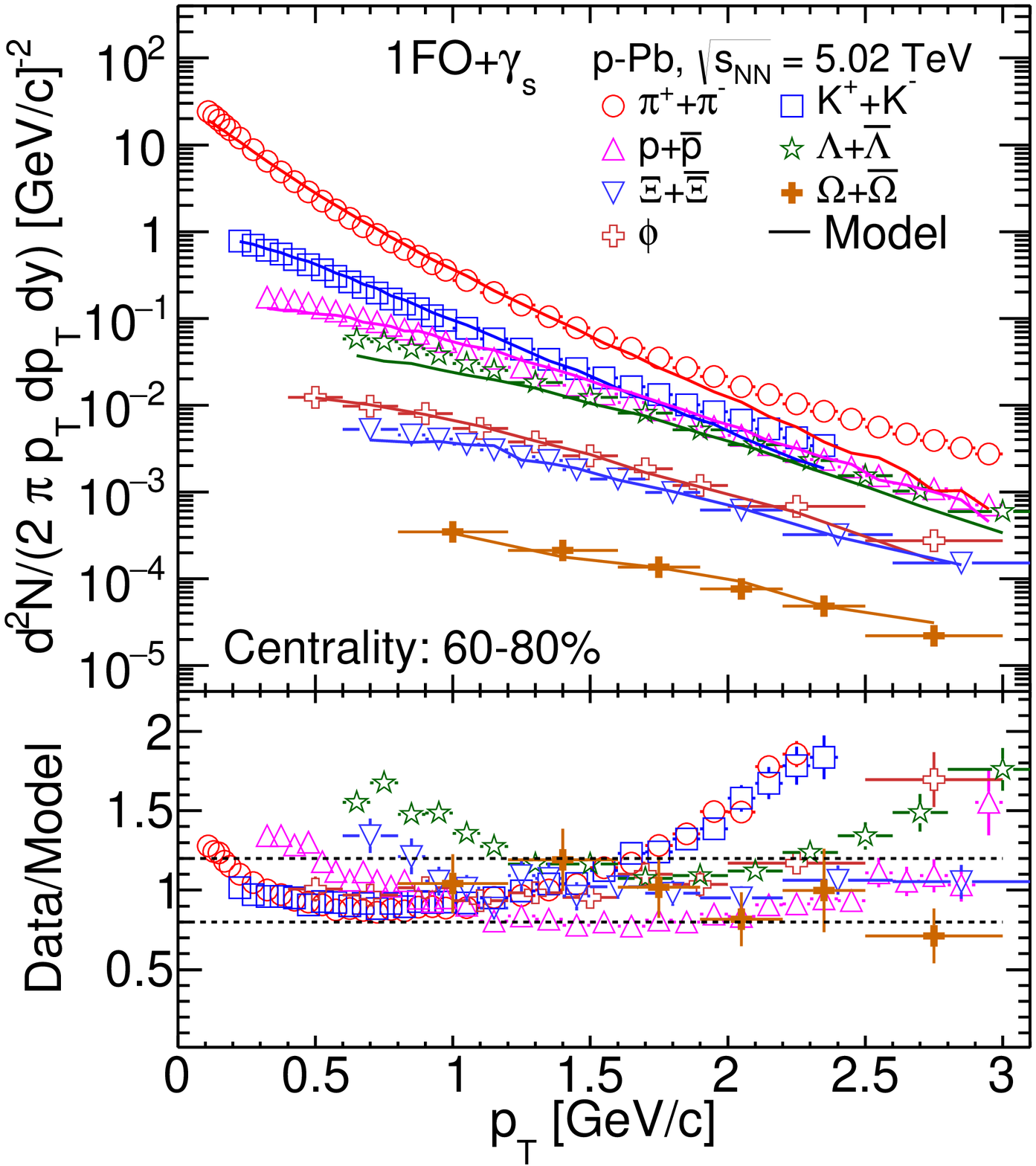} 
 \includegraphics[scale=0.28]{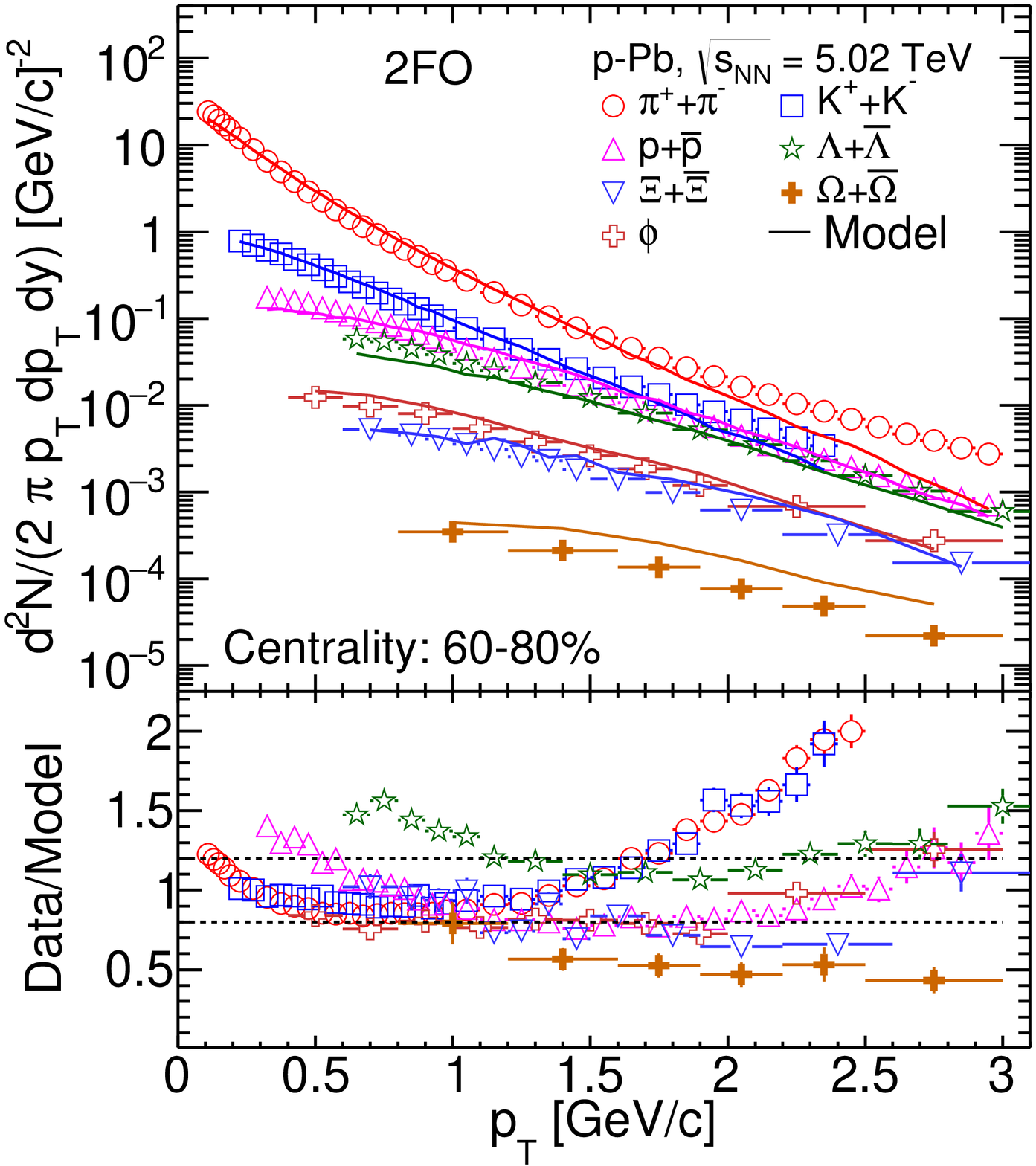}
 \caption{(Color online) The comparison of $p_T$ spectra in p+Pb collisions at $\sqrt{s_{NN}} =$ 5.02 TeV obtained from the THERMINATOR \cite{Kisiel:2005hn, Chojnacki:2011hb}  and data \cite{pPbALICEpikp, pPbALICEstrange,  pPbALICEresonance} is shown for three centralities: $0-5 \%$, $10-20 \%$ and $60-80 \%$ in three different FO schemes at $0 < y_{cm}<0.5$. The gross features of the spectra comparison seems to be independent of the freezeout scheme. The bottom panels show the ratio of data to model calculation.}
 \label{fig.spectra}
 \end{center}
\end{figure*}

 \begin{table*}[]
 \centering
 \caption{$\chi^{2}$ and $ndf$ in the 1FO, 1FO+$\gamma_{s}$ and 2FO scheme in p+Pb collisions at $\sqrt{s_{NN}}$ = 5.02 TeV.}
 \label{tab.chisqr}
 \begin{tabular}{@{}|c|c|c|c|c|c|c|c|c|c|@{}}
 \toprule
 \hline
 Centrality (\%) &  \multicolumn{3}{c}{} &\multicolumn{3}{c}{$\chi^{2}$ ($ndf$)}  & \multicolumn{3}{c|}{} \\ \cmidrule(l){2-10} 
 \cline{2-10}
   & \multicolumn{3}{c|}{1FO}                                                         & \multicolumn{3}{c|}{1FO+$\gamma_{s}$}    & \multicolumn{3}{c|}{2FO} \\ \cmidrule(l){2-10}
  \cline{2-10} 
                    & Non-Strange   & Strange &  Total             & Non-Strange & Strange &  Total      & Non-Strange & Strange &  Total      \\ \hline
 0-5             & 248 (62)          &  374(66)    &  622(131)   &  256(61) &  208(65) & 464(130)&   225(59) &  206(63) & 432(128)        \\ \hline
 5-10           & 361(62)         &  325(66)    &  686(131)     &  297(61) &  207(65) & 504(130)&   290(59) &  257(63) & 547(128)         \\ \hline
 10-20         & 377(62)         & 445(66)     &  822(131)     &  324(61) &  204(65) & 528(130)&   338(59) &  246(63) & 584(128)        \\ \hline
 20-40         & 487(62)         & 560(66)     &  1046(131)   &  392(61) &  219(65) & 612(130)&    441(59) &  320(63) & 761(128)      \\ \hline
 40-60         & 549(62)         & 822(66)     &  1371(131)   &  494(61) &  315(65) & 808(130)&    495(59) &  401(63) & 896(128)         \\ \hline
 60-80         & 716(62)         & 1463(66)     &  2179(131) &  615(61) &  251(65) & 866(130)&    671(59) &  873(63) & 1544(128)        \\ \hline
 80-100       & 824(62)         & 2428(64)     &  3253(129) &  721(61) &  311(63) & 1032(128)&     746(59) &  931(61) & 1677(126)        \\ \hline
 \end{tabular}
 \end{table*}

 \begin{table*}[]
 \centering
 \caption{Thermal freezeout parameters in the 1FO, 1FO+$\gamma_{s}$ and 2FO scheme  in p+Pb collisions at $\sqrt{s_{NN}}$ = 5.02 TeV. The average error on $T$ and $\gamma_s$ is 2 MeV and 0.02, respectively, whereas the error on $\rho_{max}$ and $\tau_{f}$ is around 15\% for all centrality classes.}
 \label{tab.fitparameter}
 \begin{tabular}{@{}|c|c|c|c|c|c|c|c|c|c|c|@{}}
 \toprule
\hline
 Centrality (\%) & \multicolumn{4} {c|} {1FO (1FO+$\gamma_{s}$)}  & \multicolumn{6}{c|}{2FO} \\ \cmidrule(l){6-11} 
\cline{6-11}
 & \multicolumn{4}{c|}{} & \multicolumn{3}{c|}{Strange}    & \multicolumn{3}{c|}{Non-Strange} \\ \cmidrule(l){2-11}
\cline{2-11} 
 & T  (MeV) & $\rho_{max}$ (fm) & $\tau_{f}$ (fm)   & $\gamma_{s}$   & T (MeV) & $\rho_{max}$ (fm)& $\tau_{f}$ (fm)& T  (MeV) & $\rho_{max}$ (fm)& $\tau_{f}$ (fm)\\ \hline
0-5             & 157(158)         &  3.9(3.8)                   &  2.7(2.7)               &   0.94                 &  160 &     3.6          &     2.5        &  154  & 4.1   &    3.0         \\ \hline
5-10           & 157(158)         &  3.5(3.5)                   &  2.6(2.6)               &   0.92                 &  160 &      3.3         &      2.3       &  154  & 3.8   &    2.8         \\ \hline
10-20         & 157(157)         & 3.3(3.4)                    &  2.5(2.5)               &   0.90                 &  160 &      3.1         &       2.2      &  154  & 3.6   &    2.7         \\ \hline
20-40         & 155(158)         & 3.1(3.0)                    &  2.4(2.4)               &   0.88                 &  158 &      2.8        &        2.2     &   152 &  3.3   &    2.7       \\ \hline
40-60         & 155(156)         & 2.7(2.7)                    &  2.2(2.3)               &   0.84                 &  157 &      2.5         &        2.0     &   153 & 2.85 &    2.35          \\ \hline
60-80         & 155(155)         & 2.2(2.3)                    &  2.0(2.1)               &   0.80                 &  156 &      2.1         &        1.9     &   153 & 2.4   &     2.2        \\ \hline
80-100       & 154(153)         & 1.6(1.7)                    &  1.9(1.9)               &   0.74                 &  155 &      1.5        &          1.7   &    153 & 1.7   &   1.9        \\ \hline
\end{tabular}
\end{table*}
	 

We have varied the $T$ in the range from 145 to 162 MeV in the steps of $1-2$ MeV whereas $\rho_{max}$ and $\tau_f$ are varied in the 
range 1.5 to 4.1 fm and 1.5 to 3.1 fm, respectively in steps of $0.1$ fm. The goodness of the parameter 
set in describing the data is ascertained from the $\chi^2/ndf$, where

\beqa
 \chi^2 &=& \sum_i \l \frac{\text{Data}\l {p_T}_i \r - \text{Model}\l {p_T}_i \r}{\text{Error}\l {p_T}_i\r} \r^2\nn\\
\eeqa
and
 $ndf$ = Number of data points - Number of free parameters.\\
 The sum goes over all available p+Pb data points up to $p_T$ = 2.5 GeV/c \cite{pPbALICEpikp, pPbALICEstrange, pPbALICEresonance}.
  For 1FO, we have varied all the three parameters $T$, $\rho_{max}$ and $\tau_f$ to 
arrive at the best parameter set. In 1FO+$\gamma_S$, we have also varied $\gamma_S$ in the range 0.7 to 1.0 in steps of 0.2 while 
for 2FO, we have varied $T$, $\rho_{max}$ and $\tau_f$ for both, non-strange as well as strange freezeout 
hypersurfaces. The $p_T$ spectra as obtained in the model for the different freezeout schemes have been 
compared with data in Fig.~\ref{fig.spectra}. In the bottom panel, we have shown the ratio of data to model. 
Unlike in Pb+Pb, where there are noticeable diagreement between 1FO and data referred to as proton anomaly 
which goes away on extending 1FO to 2FO, in p+Pb we don't find any such noteworthy tensions in 1FO. The 
quality of description of the spectra seems similar overall.

\begin{figure}[ht]
 \begin{center}
 \includegraphics[scale=0.35]{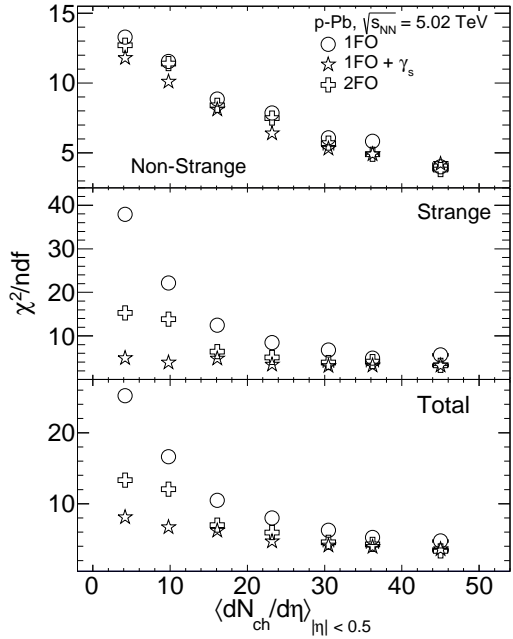}
 \caption{(Color online) The $\chi^2/ndf$ has been compared for the three different freezeout schemes across central 
 to peripheral collisions in p+Pb. 1FO$+\gamma_S$ provides the best description across all centralities. The improvement 
 over 1FO and 2FO gets better as one goes to more peripheral collisions.}
 \label{fig.chisqr}
 \end{center}
\end{figure}

The $\chi^2/ndf$ obtained in the different freezeout schemes across various centralities have been compared 
in Fig.~\ref{fig.chisqr} and the respective values of $\chi^2$ and $ndf$  are given in Table~\ref{tab.chisqr}. 
For all centralities, 1FO$+\gamma_S$ provides the 
least $\chi^2/ndf$. The improvement over 1FO and 2FO grows as one goes from central to peripheral collisions. 
This is driven by the strange sector which is more sensitive to the three different freezeout schemes studied here 
that differ in the treatment of the freezeout of strange hadrons. The yields in the non-strange sector receives 
a partial contribution from the decays of strange resonances. This leads to a small sensitivity in the fit quality 
of the non strange sector to the different freezeout schemes studied here. The improvement in 
the non-strange sector with 1FO$+\gamma_S$ is mild and uniform across centralities. We have enlisted the best parameter values that 
describe the transverse momentum spectra across different centralities within the three freezeout schemes in 
Table~\ref{tab.fitparameter}.

\begin{figure*}
 \begin{center}
 \includegraphics[scale=0.35]{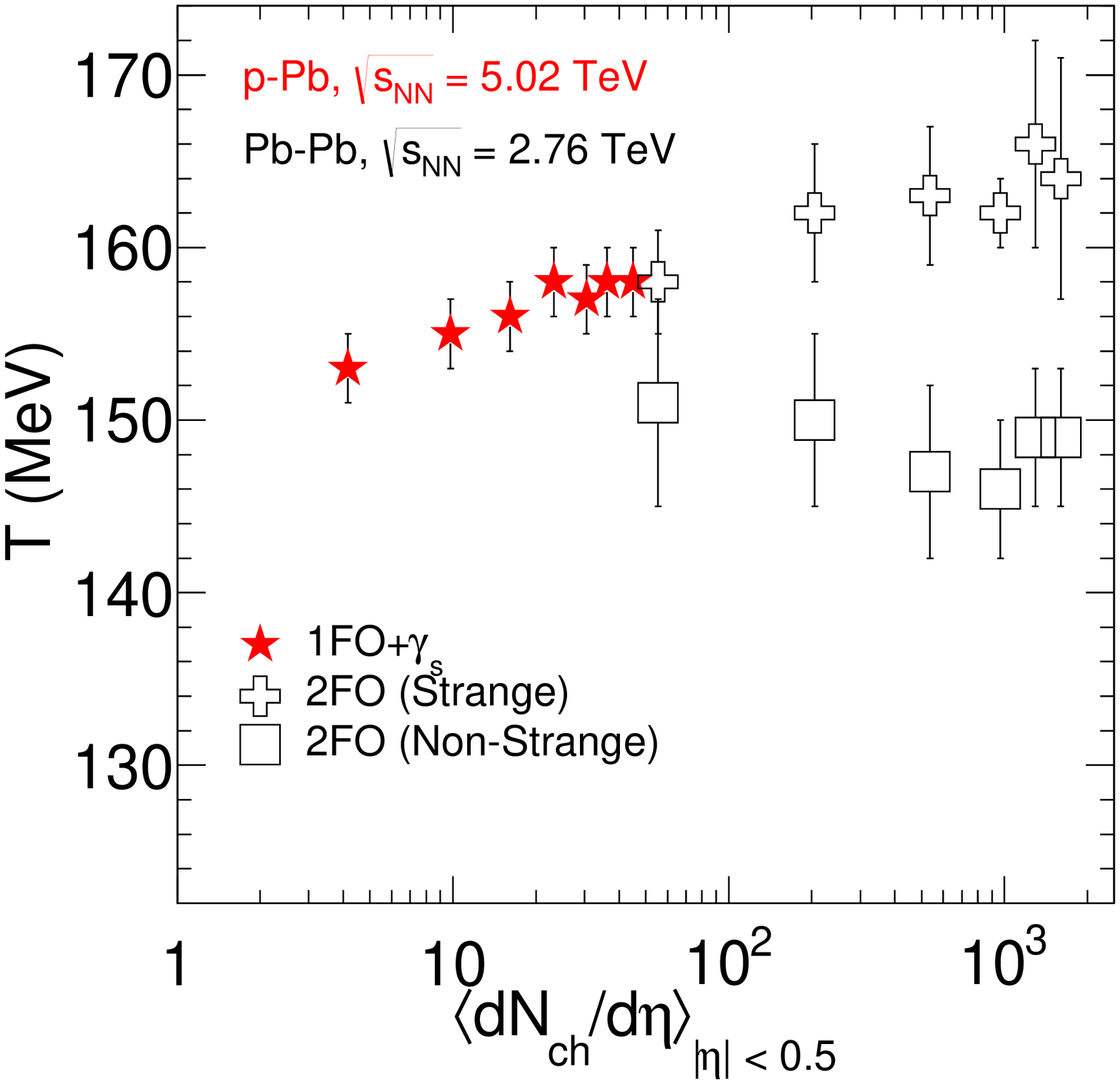}
 \includegraphics[scale=0.35]{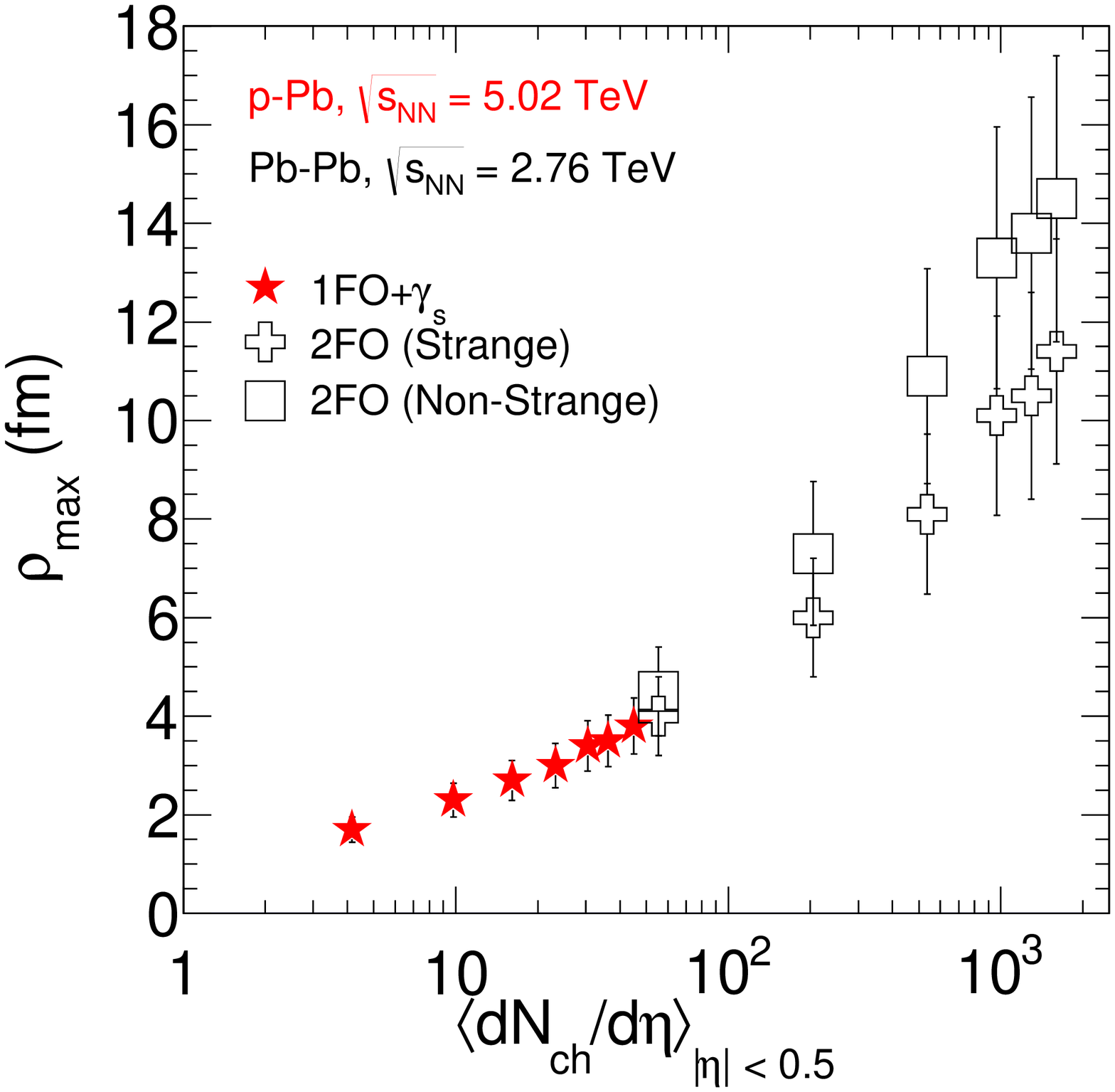}\\
 \includegraphics[scale=0.35]{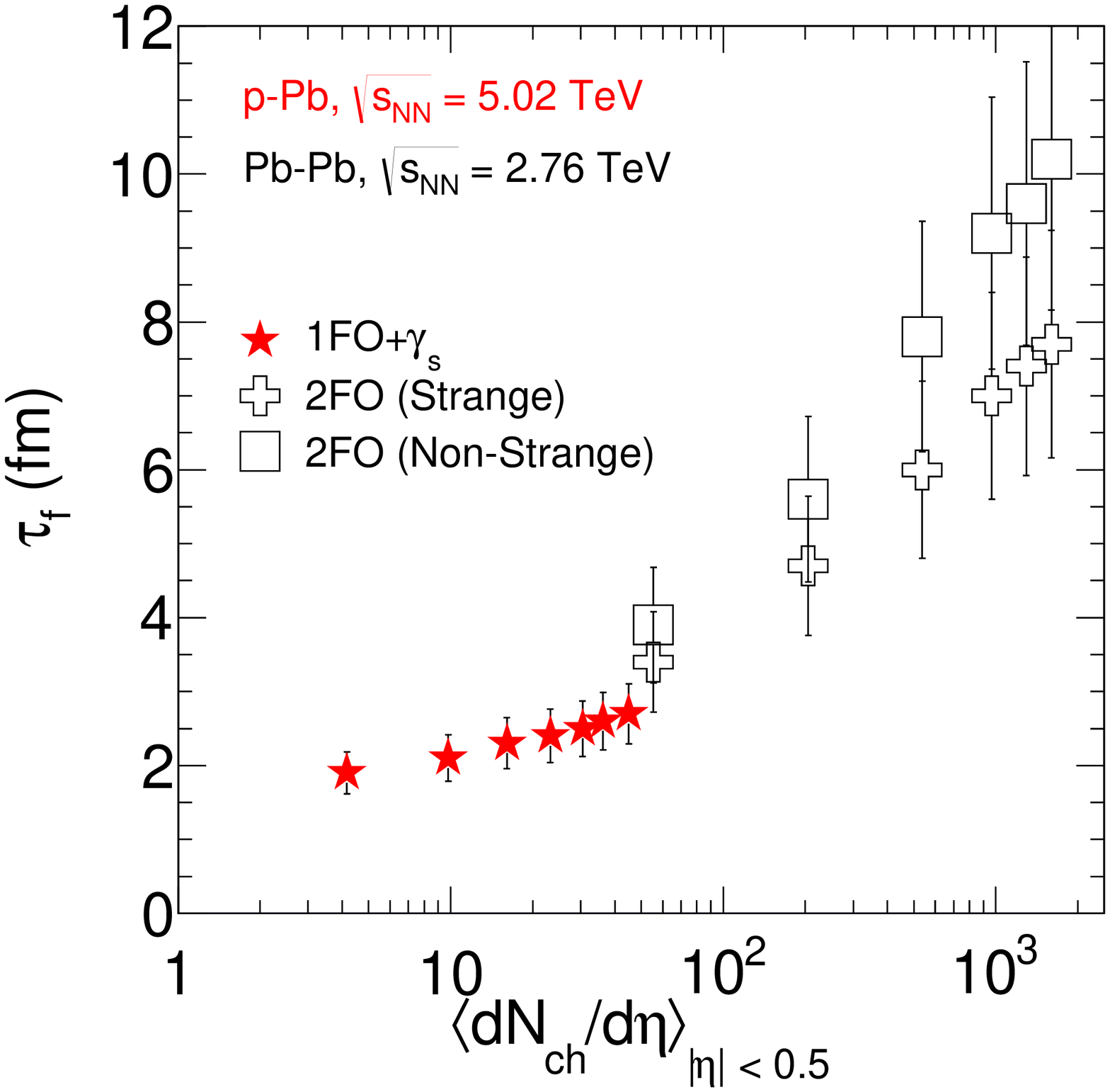}
 \includegraphics[scale=0.35]{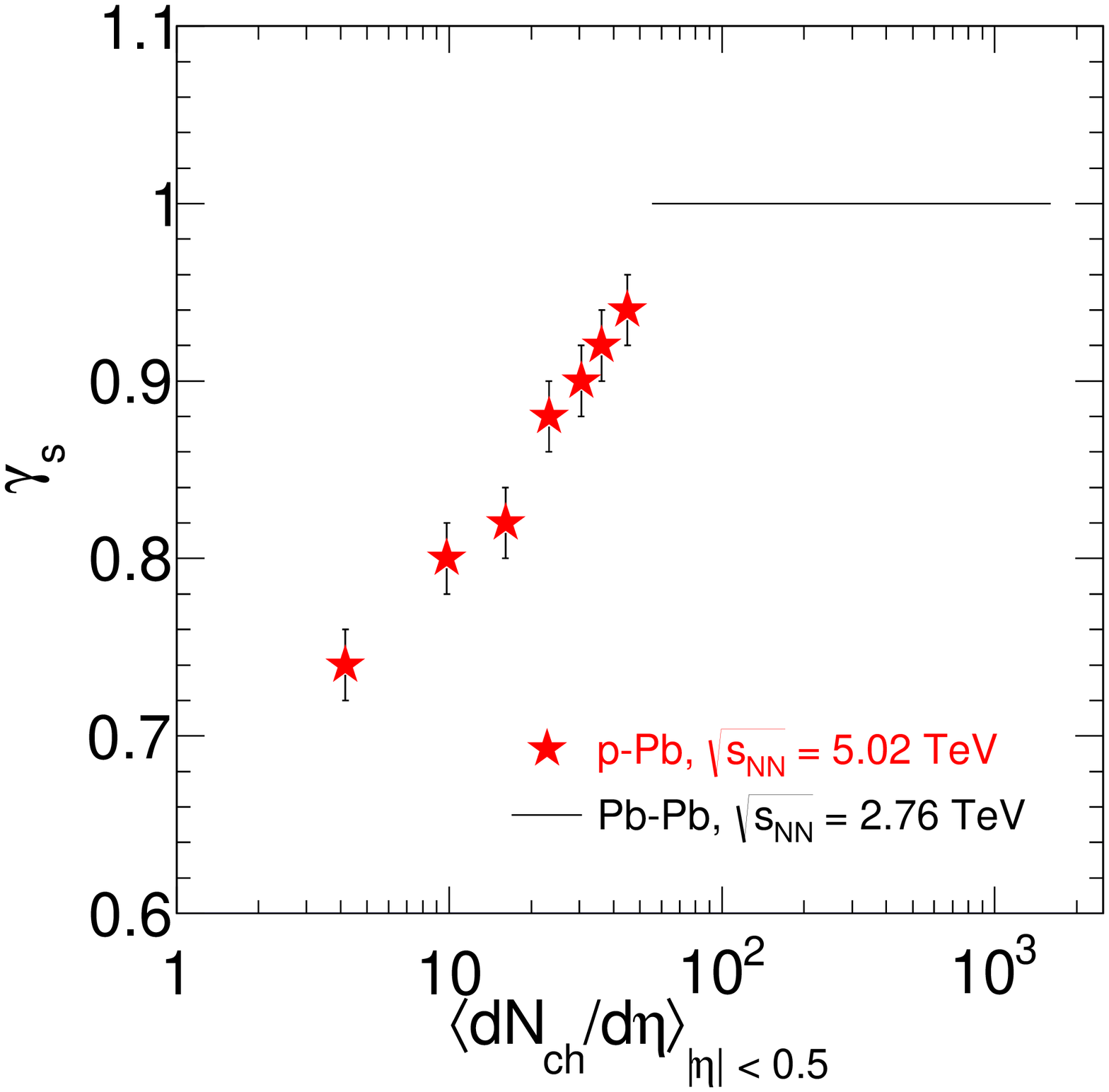} \\
 \caption{(Color online) The extracted freezeout parameters with best goodness of fit for different centralities 
 in p+Pb and Pb+Pb collisions. There is a gradual preference for sequential freezeout of the strange and non-strange 
 flavors as we go to higher multiplicity events. The $\gamma_S$ value is fixed at one for Pb+Pb collisions at $\sqrt{s}_{NN}$ = 2,76 TeV.}
 \label{fig.param}
 \end{center}
\end{figure*}

Finally in Fig.~\ref{fig.param} we have plotted the extracted freezeout parameters corresponding to the least $\chi^2/ndf$ 
with event multiplicity across different centralities in p+Pb and Pb+Pb that vary over three orders of magnitude. While the $T$ remains mostly 
flat between $145-160$ MeV, $\rho_{max}$ and $\tau_f$ show a growth of 5-7 times. The growth rate is smooth across 
system size. We also note that the difference between the non-strange and strange freezeout parameters systematically 
increase as we go to events with higher multiplicity, signifying the role of interaction. However, currently the uncertainties 
over the extracted parameters in the non-strange and strange sectors are large and does not allow us to futher quantify the 
magnitude of the hierarchy in freezeout of the strange and non-strange flavors. $\gamma_S$ in p+Pb steadily grows from 
0.74 to about 0.94 across peripheral to central collisions. The approach to strangeness equilibration with more central 
p+Pb events could be related to the larger entropy deposition in the initial state in central p+Pb collisions as opposed 
to peripheral events~\cite{Castorina:2017dvt}. We use similar errors on T, $\rho_{max}$ and $\tau_f$ as for Pb+Pb results \cite{Chatterjee:2014lfa} since the errors are mostly system size independent.


\section{Summary and Outlook}
\label{sum}
The hadron yields and $p_T$ spectra are the standard observables to throw light on freezeout dynamics. Contrary to 
expectations, the 1FO scheme is known to be blind to system size dependence in freezeout~\cite{Becattini:2010sk}. 
However, simultaneous analysis of the hadron yields in Pb+Pb, p+Pb and p+p revealed an interesting system size 
dependence of the preferred freezeout scheme- 2FO is preferred over 1FO and 1FO$+\gamma_S$ in Pb+Pb while in small 
systems like p+Pb and p+p, 1FO$+\gamma_S$ is preferred~\cite{Chatterjee:2016cog}. In order to put this hypothesis on 
a more strong footing, here we extend the previous analysis to hadron spectra. 
While 2FO is known to describe the hadron spectra better in Pb+Pb, here we analyse the data for different centralities in 
p+Pb. We find that allowing for a different hypersurface for the freezeout of the strange hadrons do not improve the 
quality of the fits. This is in accordance to our previous study with the hadron yields~\cite{Chatterjee:2016cog}. Thus, our current analysis 
with the data on hadron spectra reaffirms the hypothesis on the system size dependence of freezeout scheme: flavor dependent 
freezeout scheme is preferred in large systems while unified freezeout is preferred in small systems. Thus, the role of 
interaction in larger system is mostly to delay the freezeout of the non-strange hadrons.\\

\section{Acknowledgement}
AKD and RS acknowledge the support of XIIth plan project no. 12-R$\&$D-NIS-5.11-0300 of Govt. of India.
BM acknowledges financial support from J C Bose National Fellowship of DST Govt. of India.
SC acknowledges the support of the AGH UST statutory tasks No. 11.11.220.01/1 within subsidy of the Ministry of Science 
and Higher Educations and by the National Science Centre Grant No. 2015/17/B/ST2/00101.
\bibliographystyle{apsrev4-1}
\bibliography{SystemSizeSpectra}

\end{document}